\documentclass[a4paper,11pt]{article}

\usepackage[english]{babel}
\usepackage[utf8]{inputenc}
\usepackage{amsmath}
\usepackage{amsfonts}
\usepackage{graphicx}
\usepackage{authblk}
\usepackage[colorinlistoftodos]{todonotes}

\usepackage[inner=3.0cm,outer=2.5cm,top=2.5cm,bottom=2.5cm,includeheadfoot]{geometry}

\usepackage{setspace}
\title{Short pulse generation and mode control of broadband terahertz quantum cascade lasers}

\author[1,$\dagger$,*]{Dominic Bachmann}
\author[2,$\dagger$,*]{Markus R\"osch}
\author[2]{Martin J. S\"uess}
\author[2]{Mattias Beck}
\author[1]{Karl Unterrainer}
\author[1]{Juraj Darmo}
\author[2]{J\'er\^ome Faist}
\author[2]{Giacomo Scalari}

\affil[1]{Photonics Institute and Center for Micro- and Nanostructures, TU Wien, Gu{\ss}hausstra{\ss}e 27-29, 1040 Vienna, Austria}
\affil[2]{Institute of Quantum Electronics, ETH Z\"urich, Auguste-Piccard-Hof 1, 8093 Zurich, Switzerland}
\affil[$\dagger$]{These authors contributed equally to this work}
\affil[*]{Corresponding authors: dominic.bachmann@tuwien.ac.at, mroesch@phys.ethz.ch}

\date{}

\begin{document}
	
\maketitle

\begin{abstract}
	We report on a waveguide engineering technique that enables the generation of a bandwidth up to $\sim$\,1\,THz and record ultra-short pulse length of 2.5\,ps in injection seeded terahertz quantum cascade lasers. The reported technique is able to control and fully suppress higher order lateral modes in broadband terahertz quantum cascade lasers by introducing side-absorbers to metal-metal waveguides. The side-absorbers consist of a top metalization set-back with respect to the laser ridge and an additional lossy metal layer. In continuous wave operation the side-absorbers lead to octave spanning laser emission, ranging from 1.63 to 3.37\,THz, exhibiting a 725\,GHz wide flat top within a 10\,dB intensity range as well as frequency comb operation with a bandwidth of 442\,GHz. Numerical and experimental studies have been performed to optimize the impact of the side-absorbers on the emission properties and to determine the required increase of waveguide losses. Furthermore, these studies have led to a better understanding of the pulse formation dynamics of injection-seeded quantum cascade lasers.
\end{abstract}

\newpage
\section{Introduction}
In recent years, it was demonstrated that quantum cascade lasers (QCLs) \cite{faist1994} are very well suited to develop broadband sources of coherent radiation in the terahertz (THz) spectral region \cite{turcinkova2011,Roesch2014}. The concept of heterogeneous QCLs, firstly introduced in the mid-infrared \cite{Gmachl2002}, has proven to be very powerful in the THz region as well. It relies on stacking different active regions with individually designed emission frequencies into a single waveguide. Such heterogeneous active regions offer high design freedom to shape the optical gain and thereby allow achieving lasing bandwidths well beyond 1\,THz \cite{turcinkova2011,Roesch2014}. Furthermore, it was shown that the group velocity dispersion (GVD) in such devices is sufficiently low so they can operate as frequency combs (FCs) without any dispersion compensation \cite{Roesch2014,Roesch2016}. Applications for compact and efficient FCs based on THz QCLs are manifold and range from spectroscopy to metrology. However, non-uniform power distribution of the individual lasing modes, a limited spectral bandwidth and higher order lateral modes often hamper the performance of such lasers, especially in continuous wave operation \cite{turcinkova2011,Burghoff2014}.

Another interesting field of use for heterogeneous THz QC gain media is the development of broadband amplifiers \cite{bachmann2015}, the coherent detection of its emitted radiation \cite{oustinov2010,bachmann2015} and the ultra-short pulse generation by mode-locking \cite{barbieri2011,Freeman2013,Wang2015}. In contrast to previous contributions, which all rely on homogeneous active regions, heterogeneous lasers potentially allow the generation of sub-ps long THz pulses with bandwidth exceeding 1\,THz.

The metal-metal waveguide \cite{williams2003} is the resonator of choice for all broadband concepts mentioned above. Its essential properties include the close to unity confinement of the optical mode and an almost homogeneous field distribution across the active region \cite{scalari2008review,kohen2005}. Additionally, and particularly important for FC operation, it has nearly constant waveguide losses and group velocity dispersion between 2 and 4\,THz \cite{Roesch2014}. These advantages come along with some undesired properties such as the significantly lower optical output powers \cite{williams2007} and the more divergent beam patterns \cite{Adam2006,Orlova2006} compared to lasers with single plasmon waveguides\,\cite{KOE02}.

Besides the mentioned extraction problem, another major issue of metal-metal waveguides are higher order lateral modes that are not fully suppressed. Such modes are undesired for short pulse generation and also hinder FC operation. The generation of short pulses in broadband THz QC gain media is up to date limited by the formation of sub-pulse structures, favored by the short upper state lifetime of the gain medium, that hinder regular pulse train formation \cite{bachmann2015}. So far, the origin of this sub-pulse structure has not yet been fully understood. As the spectra of such time traces indicate the presence of higher order lateral modes, they might be the source of this sub-pulse formation. Higher order lateral modes do not encounter the same effective refractive index as the fundamental modes, resulting in different group velocities and the build up of sub-pulses. A successful suppression of higher order lateral modes is therefore of paramount importance in order to generate ultra-short pulses down to a ps level.

Furthermore, higher order lasing modes cause a very unevenly distributed power over the fundamental cavity modes due to mode competition. Even though the modal overlap of the excited lateral modes with the active region is reduced for narrow ridges with widths of the order of 50\,$\mu$m \cite{Maineult2008}, it still does not completely hinder them from lasing. An elegant way to achieve a full suppression is to increase the difference in losses between the fundamental and the higher order lateral modes by implementing side-absorbers to the metal-metal waveguide \cite{Fan2008,xu2013,Tanvir2013}. This can be done by slightly reducing the top metalization width with respect to the laser ridge width (set-back) and an underlying highly doped, and therefore very lossy, epitactically grown GaAs layer. 

In this article, we present a very simple but versatile way to control and fully suppress higher order lateral modes in broadband metal-metal THz QCLs and THz QC amplifiers. In contrast to the work reported in Ref. \cite{Fan2008,xu2013,Tanvir2013,Gellie2008} we introduce a set-back along with additional losses given by a thin metallic layer, which is deposited onto functional devices, instead of using a highly-doped epitaxial layer. Finite element simulations are used to obtain frequency dependent optical losses in order to choose the right set-back width with respect to the laser waveguide width. The correct implementation eliminates mode competition effects and thus tremendously improves the lasing bandwidth and the power distribution between individual modes. Furthermore, it ensures that only fundamental cavity modes are excited by injection seeding with broadband THz pulses. As a consequence, a clean train of transform-limited pulses is formed, achieving record short pulse widths.

\section{Lateral mode control}
\label{sec:sim}
\begin{figure}[h!]
	\centering
	\includegraphics[width=0.7\linewidth]{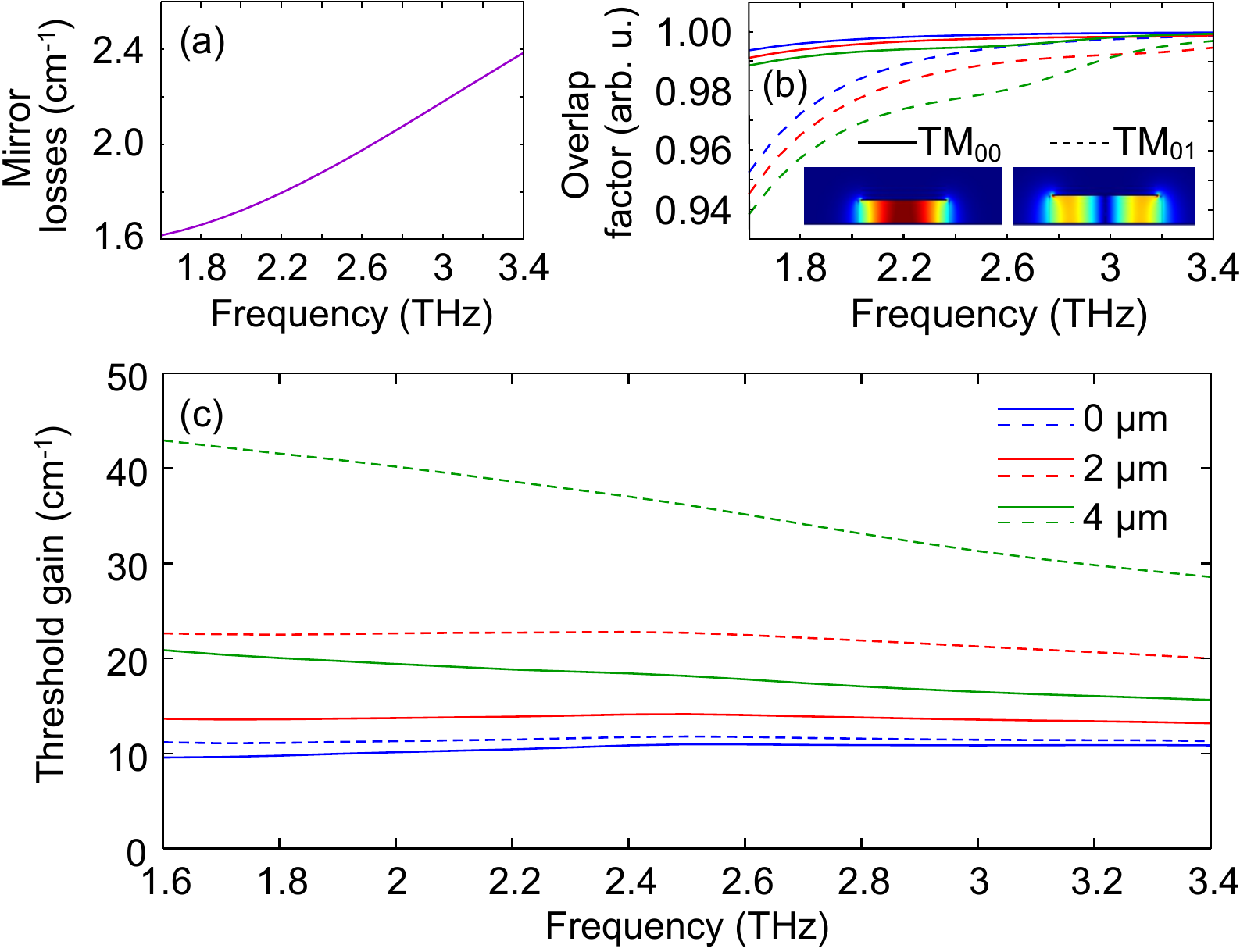}
	\caption{Simulations. (a) Mirror losses of a 50\,$\mu$m wide, 13\,$\mu$m high and 2\,mm long laser. (b) Overlap factor for the same ridge with different amounts of lossy side-absorbers (blue:\,none, red:\,2\,$\mu$m, green:\,4\,$\mu$m). Solid lines are calculations for $\text{TM}_{00}$ modes and dashed lines for $\text{TM}_{01}$ modes. The inset displays the electric field distribution of these two modes in a 13\,$\mu$m high waveguide. (c) Impact of lossy side-absorbers (set-back with 5 nm of nickel on top) on the threshold gain for a 50\,$\mu$m wide ridge.}
	\label{fig:sim1}
\end{figure}

To fully understand the impact of lossy side-absorbers on QC structures, threshold gain calculations were performed. In contrast to the models reported in Ref.\,\cite{Fan2008,xu2013}, we performed frequency dependent simulations in order to adapt them to broadband lasers. The threshold gain $g_{th}$ is defined as \cite{faist2013book} 

\begin{equation}
g_{th}(\nu)=\frac{\alpha_m(\nu)+\alpha_w(\nu)}{\Gamma(\nu)},
\label{eq:threshold}
\end{equation}
\\
where $\alpha_m$ and $\alpha_w$ are the mirror and waveguide losses of the laser cavity. The mirror losses have been evaluated for a 2\,mm long and 13\,$\mu$m high laser with different ridge widths using a 3D finite element solver (\textsc{CST Microwave Studio$^{\circledR}$}). The mirror losses of a 50\,$\mu$m wide ridge are displayed in Fig.\,\ref{fig:sim1}(a). For simplicity we were using the same mirror losses for the first ($\text{TM}_{00}$) and second ($\text{TM}_{01}$) lateral mode in the waveguide. $\Gamma$\,in \eqref{eq:threshold} represents the overlap factor of the electric field with the active region and is defined as the ratio of the amount of optical intensity inside the active region normalized to the total intensity \cite{faist2013book}

\begin{equation}
\Gamma(\nu)=\frac{\int_{active}dzdxE_z^2(z,x,\nu)\epsilon(z,x,\nu)}{\int_{z,x=-\infty}^{\infty}dzdxE^2(z,x,\nu)\epsilon(z,x,\nu)},
\end{equation}
\\
where $E(z,x,\nu)$ is the electric field at position $(z,x)$ with a frequency $\nu$, $E_z(z,x,\nu)$ is the part of the electric field coupling to the intersubband transition and $\epsilon(z,x,\nu)$ is the dielectric function at the frequency $\nu$ for the material at position $(z,x)$. $\Gamma$ has been calculated in a 2D simulation using the finite element solver \textsc{Comsol Multiphysics$^{\circledR}$} (Fig.\,\ref{fig:sim1}(b)). The dielectric function of GaAs has been approximated using a Lorentz oscillator model. The calculations have been performed for the $\text{TM}_{00}$ (solid lines) as well as for the $\text{TM}_{01}$ modes (dashed line). The corresponding electric field distributions of these modes inside the waveguide are shown in the inset of Fig.\,\ref{fig:sim1}(b). Different waveguide configurations have been simulated. For the blue curves in Fig.\,\ref{fig:sim1}(b) the top metalization has the same width as the active region below, i.e. there are no lossy side-absorbers. For the other two configurations, the width of the top metalization is reduced on both sides by 2\,$\mu$m/4\,$\mu$m (red/green lines). In addition, in order to get lossy side-absorbers, 5\,nm of nickel has been added on top of these set-backs. As can be seen in Fig.\,\ref{fig:sim1}(b), the side-absorbers change the overlap factor only by a small amount. For both, $\text{TM}_{00}$ and $\text{TM}_{01}$ modes, $\Gamma$ is close to unity for all investigated configurations.

The same 2D simulations have been used to calculate the complex refractive index $\tilde{n}(\nu)=n-ik$ of the modes propagating in such a waveguide. The extinction coefficient $k$ is directly linked to the waveguide losses by \cite{saleh_teich_book}

\begin{equation}
\alpha_w(\nu)=\frac{4\pi}{\lambda}k(\nu).
\end{equation}
\\
These losses have been inserted into \eqref{eq:threshold} together with the intersubband losses of the active region as reported in Ref. \cite{Roesch2014} to calculate the threshold gain. As shown in Fig.\,\ref{fig:sim1}(c), for a 50 $\mu$m wide waveguide without set-back, the $\text{TM}_{00}$ and the $\text{TM}_{01}$ modes have an almost identical threshold gain over the entire frequency range of the laser (blue curves). It is therefore not surprising that in such a device not only the fundamental modes reach lasing threshold. The resulting gain competition will lead to a very inhomogeneous laser spectrum. If one adds a set-back of 2\,$\mu$m per side with 5\,nm of nickel on top, the threshold gain for the $\text{TM}_{01}$ mode is increased up to 22 $\text{cm}^{-1}$ while the $\text{TM}_{00}$ mode only increases slightly to approximately 14 $\text{cm}^{-1}$ (red curves). As a consequence, only the fundamental modes will reach lasing threshold. This results in a much more homogeneous power distribution and a clearly defined mode spacing, which is defined by the cavity round-trip frequency, as there are no more $\text{TM}_{01}$ modes and thus no gain competition. But, if the set-back is chosen too wide, only part of the laser bandwidth will reach threshold (green curves in Fig.\,\ref{fig:sim1}(c)), and an even further increase of the threshold gain will result in no lasing at all.

The waveguide losses are strongly dependent on the width of the laser. Since the overlap of the electric field with the lossy-absorbers is smaller for wider ridges, they require a broader set-back in order to achieve the same shift in threshold gain, compared to narrow ridges. From our simulations, we get an optimal trade-off between the threshold gain of $\text{TM}_{00}$ and $\text{TM}_{01}$ as follows: For 50\,-\,70\,$\mu$m wide ridges a set-back of 2\,-\,4\,$\mu$m per side is optimal, whereas for wider ridges of 100\,-\,150\,$\mu$m, a set-back of 10\,-\,12\,$\mu$m is required.

\section{Sample fabrication and characterization}
\label{sec:fabrication}
The sample fabrication is based on a standard metal-metal processing technique that relies on a thermo-compression wafer bonding step of the 13\,$\mu$m thick active region onto a highly doped GaAs substrate \cite{williams2003}. The employed heterogeneous active region is identical to the one presented in Ref. \cite{Roesch2014} and consists of three different GaAs/AlGaAs quantum cascade heterostructures. These are designed for center frequencies of 2.3, 2.6 and 2.9 THz and the total structure is able to generate octave spanning laser emission\,\cite{Roesch2014}. The highly doped GaAs top contact layer was removed and the laser cavities were defined by a reactive ion etching (RIE) process, resulting in almost vertical side walls. Thus, the devices do not possess a significant amount of unpumped active region, which is beneficial for the thermal management, particularly relevant for continuous wave operation. Additionally, dry etched resonators are more suitable for long (2-4\,mm) and narrow (40-80\,$\mu$m) lasers. Those feature a larger number of fundamental Fabry-Perot modes, important for spectroscopy applications, while reducing the probability of higher order lateral modes reaching laser threshold.

The lossy side-absorbers were implemented on devices for amplification and pulse generation. The geometry is based on a coupled cavity system that is designed for THz time-domain spectroscopy experiments, where a short section acts as an integrated emitter of broadband THz pulses and a long section as an amplifier for the injected THz radiation \cite{Martl2011,Bachmann2014}. The 3\,$\mu$m wide (each side) set-back of the top metal contact layer (Ti/Au) with respect to the total laser ridge width of 75\,$\mu$m was defined by a sacrificial silicon nitride (SiNx) etch mask, used on the amplifier section only. As a final step, the required amount of side-absorber loss is introduced by evaporating a thin layer (5\,nm) of nickel on top of the fully processed device. The choice of appropriate set-back width and nickel thickness, for a given resonator width, is the crucial point for good spectral performance, as discussed in section \ref{sec:sim}. Figure \ref{fig:sem-cc}(a) shows a scanning electron microscope (SEM) image of such a coupled cavity device.

In order to better understand the effect of side-absorbers on the spectral performance as well as the effects on FC operation, a second device geometry has been fabricated. It consists of a single QCL ridge and the lossy side-absorbers were introduced as a post-processing step. The set-back was realized by removing a stripe of the top metalization, on either one or both waveguide edges, by a focused ion beam (FIB; System: Helios 600i). As for the previous geometry, losses were induced by evaporating 5\,nm of nickel onto the region of the ridge now uncovered. This approach is very powerful, because it allows to adapt the set-back width on demand. Furthermore, it permits a "before and after" comparison on the very same laser, directly revealing the impact of this mode control scheme. The technique of FIB cutting can also be used for realizing integrated planar horn structures in order to improve the outcoupling efficiency and the far-field pattern of metal-metal THz QCLs, as it was shown in \cite{wang2016}. 

Figure \ref{fig:sem-cc}(b) displays the light-current-voltage characteristics (LIV) of a 2\,mm x 50\,$\mu$m dry etched laser with (red line) and without (blue line) a lossy side-absorber, consisting of a 3\,$\mu$m wide set-back on one side and 5\,nm of nickel. The measurements were performed in pulsed operation at a heat-sink temperature of 10\,K. Consistent with the simulations, we observe a slight increase of the threshold current density as a result of the increased waveguide losses due to the FIB/evaporation treatment. Additionally, the THz output power is about 30\% higher, which we mainly attribute to an improved collection efficiency, resulting from a more directed far-field since the higher order lateral modes are suppressed. This is discussed in more detail in the following section.

\section{Laser spectrum and far-field pattern}
\begin{figure}[htbp]
	\centering
	\includegraphics[width=0.7\linewidth]{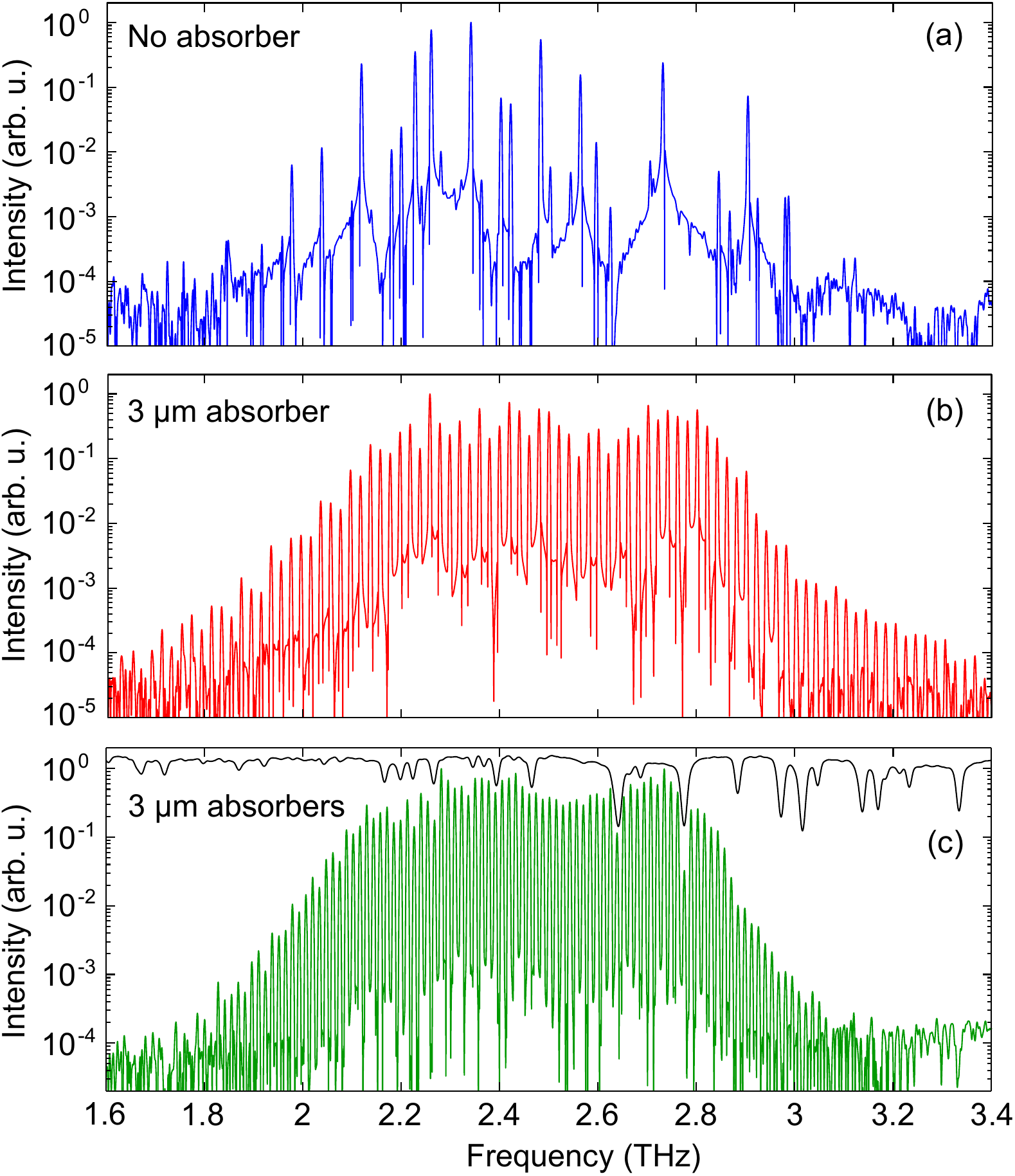}
	\caption{Suppression of higher order lateral modes by lossy side-absorbers and its impact on the laser spectrum of broadband QCLs. (a) Typical spectrum of a 2\,mm long and 70\,$\mu$m wide dry etched laser without set-back. (b) Laser spectrum of the same QCL after removing a stripe of 3\,$\mu$m on one side of the top metalization with a FIB and the deposition of 5\,nm nickel. The device shows an octave-spanning bandwidth. (c) Spectrum of a coupled cavity QCL device with a 2.93\,mm long and 75\,$\mu$m wide laser section, including lossy side-absorbers on both sides. The black line represents the atmospheric absorption of the used non-vacuum FTIR system (not to scale).}
	\label{fig:spectrum-setback}
\end{figure}

While for wet etched lasers a set-back of the top metalization is routinely implemented to prevent underetching, this is not necessary for dry etching, as a good etch conditioning will result in almost vertical side walls. A laser spectrum of a typical dry etched device (2\,mm\,x\,70\,$\mu$m) without any set-back is displayed in Fig.\,\ref{fig:spectrum-setback}(a). It is apparent that such a laser cannot be used as a frequency comb source. It has only few modes with an irregular spacing and the power is distributed very unevenly between them. To directly show the effect of side-absorbers on the very same laser, a set-back of 3\,$\mu$m (on one side only) has been removed from the top metal using a FIB. After depositing additional 5\,nm of nickel, the laser emission was measured again. The spectrum with the implemented side-absorber is shown in Fig.\,\ref{fig:spectrum-setback}(b). Not only the lasing of higher order modes is suppressed, but also the overall shape of the spectrum is significantly improved. The presence of the side-absorber ensures a smooth mode intensity envelope, leading to a bandwidth of 720\,GHz within a 10\,dB limit. Furthermore, the total spectral bandwidth increases, demonstrating octave spanning laser emission ranging from 1.63 to 3.37\,THz. As already shown in Ref.\,\cite{Roesch2014}, the weak modes on the edges of the spectrum are indeed lasing modes and are by no means measurement artifacts. The laser spectra in Fig.\,\ref{fig:spectrum-setback}(a) and (b) were measured in continuous wave operation at a heat-sink temperature of 20\,K. Fig.\,\ref{fig:spectrum-setback}(c) shows the effect of side-absorbers in pulsed operation on a device with emitter section, as discussed later.

\begin{figure}[htbp]
	\centering
	\includegraphics[width=0.7\linewidth]{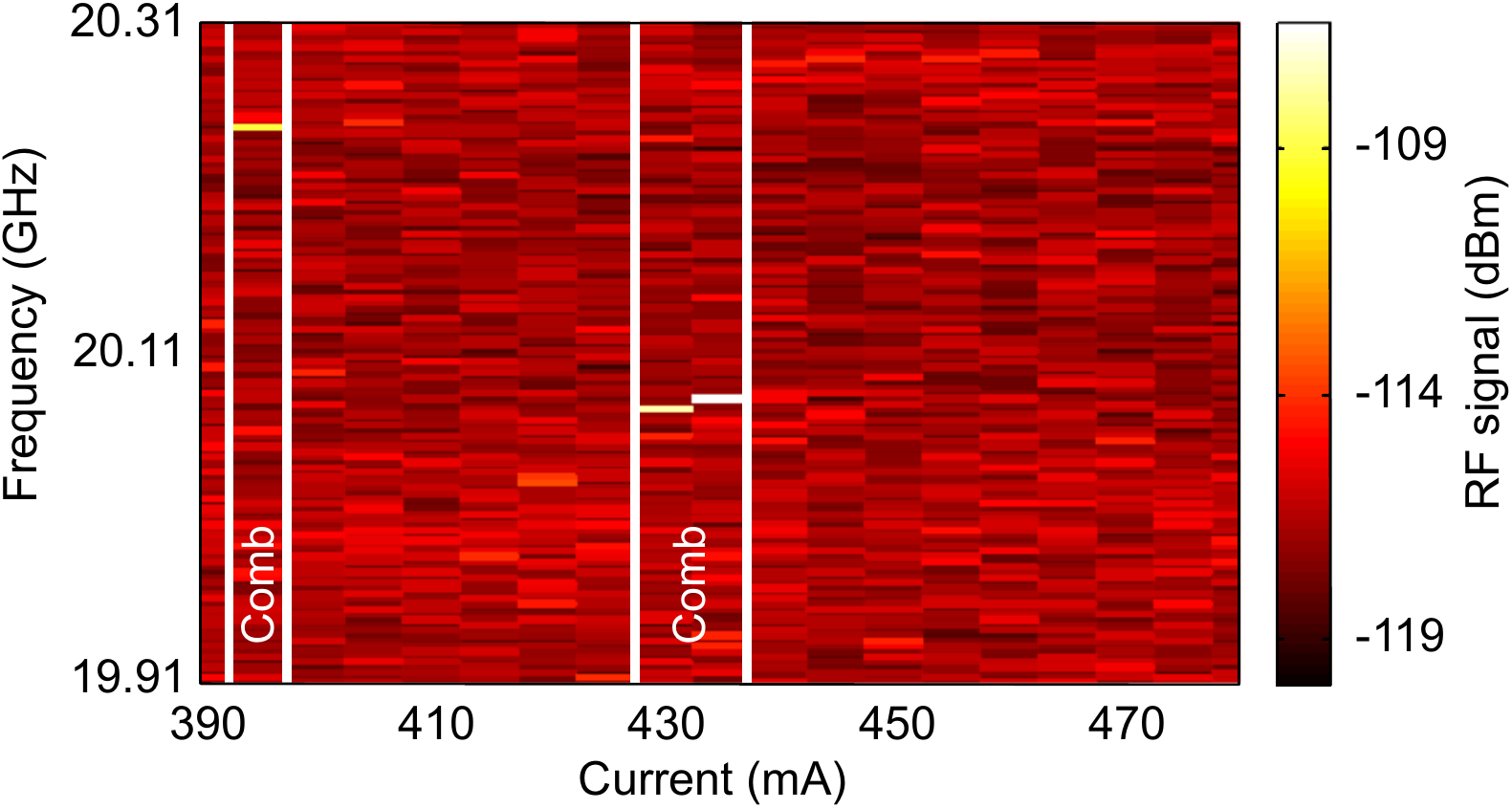}
	\caption{Electrical beatnote as a function of laser bias current, measured in CW at 20\,K. The beatnote indicates two FC regimes centered at 395\,mA and at 430\,mA, visible as narrow RF peaks.}
	\label{fig:BNmap}
\end{figure}

\newpage
Analyzing the electrical intermode beatnote of the laser without emitter after the implementation of side-absorbers in Fig.\,\ref{fig:BNmap} shows two regimes with narrow beatnotes, indicating FC operation of the laser \cite{hugi2012,Roesch2014}. The corresponding spectral bandwidth expands up to 442\,GHz (see supplementary data for details). The side-absorbers do not only improve the spectral properties but also promote the generation of FCs. A more detailed study of the ridge width dependence on the side-absorbers can be found in the supplementary data.

\begin{figure}[h!]
	\centering
	\includegraphics[width=0.7\linewidth]{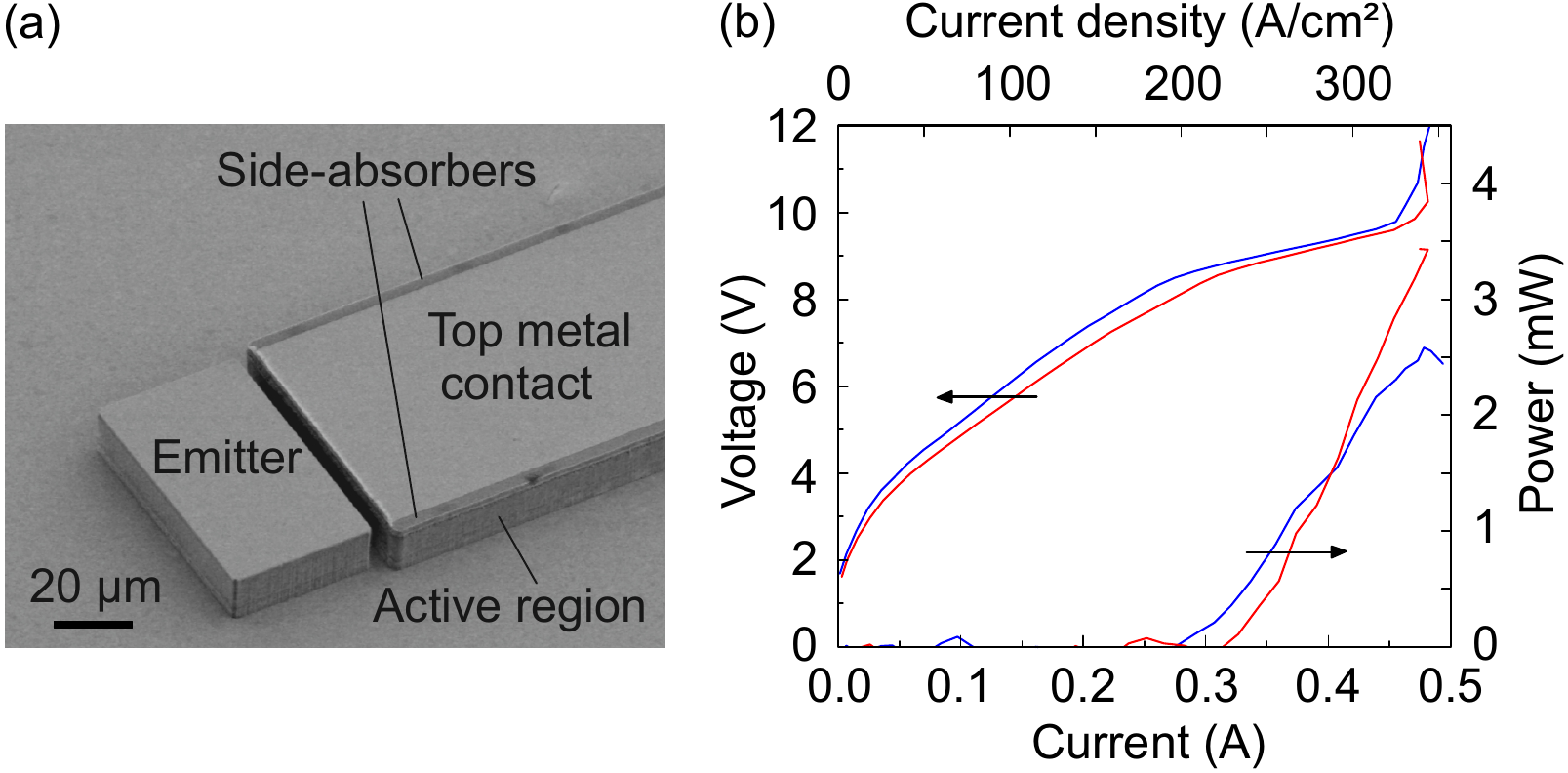}
	\caption{(a) Scanning electron microscope image of a coupled cavity THz QCL device, processed with lossy side-absorbers (3\,$\mu$m wide set-back and 5\,nm of nickel) on the long amplification section only. (b) Light-current-voltage characteristics of a 2\,mm\,x\,70\,$\mu$m dry etched metal-metal THz QCL, measured before (blue line) and after (red line) introducing a lossy side-absorber on one side.}
	\label{fig:sem-cc}
\end{figure}

\begin{figure}[htbp]
	\centering
	\includegraphics[width=0.7\linewidth]{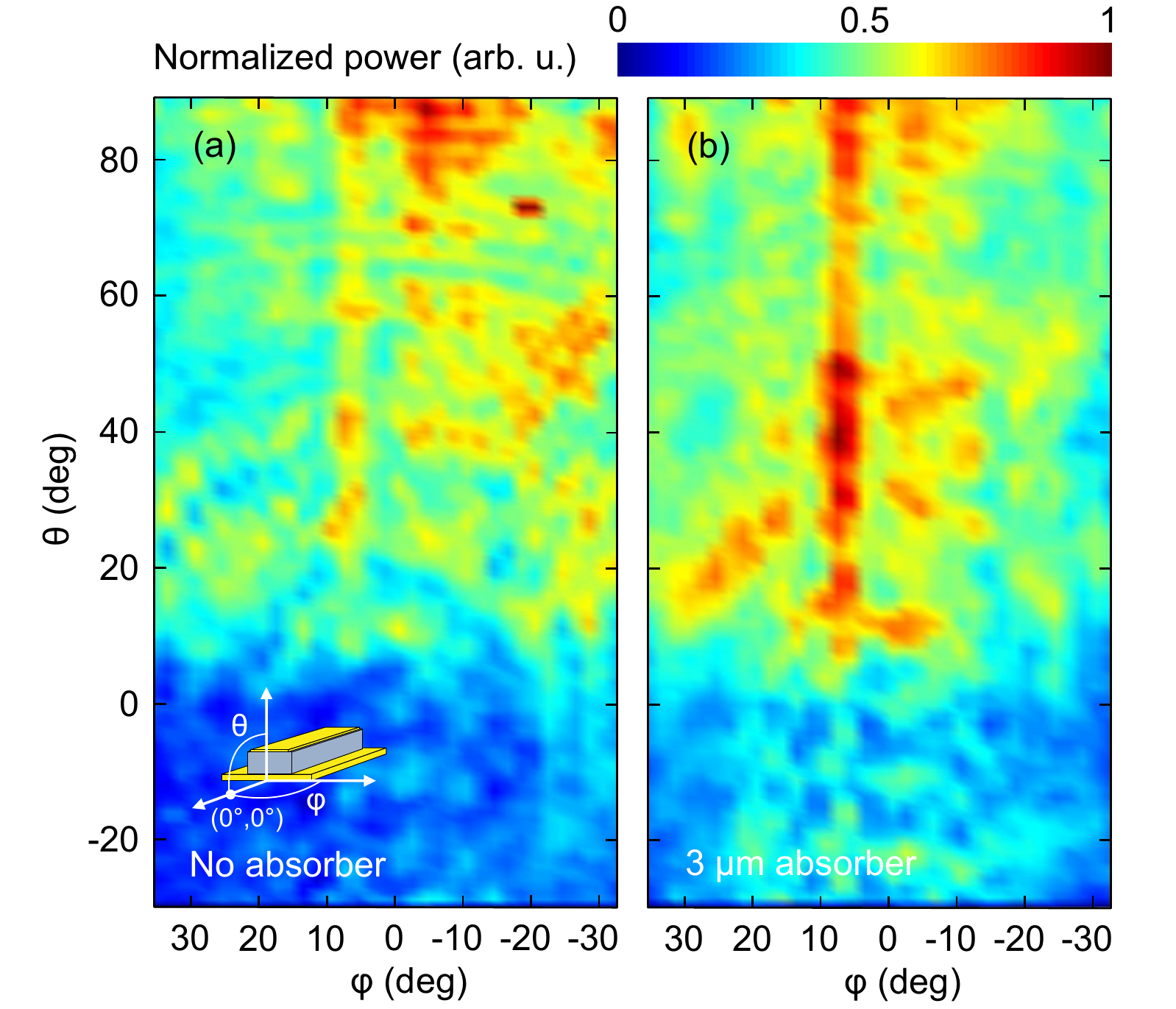}
	\caption{Far-field pattern of a 2\,mm long and 70\,$\mu$m wide laser before (a) and after (b) implementing a 3\,$\mu$m wide side-absorber in pulsed operation. Due to the suppression of $\text{TM}_{01}$ modes, the far-field in (b) is more directional in the $\phi$ axis.}
	\label{fig:farfield}
\end{figure}

We also measure more optical power when the side-absorber is present, as shown in the LIV comparison in Fig.\,\ref{fig:sem-cc}(b). This increase in power can be attributed to a change of the far-field. Even though the beam patterns of standard metal-metal waveguides are highly non-directional due to their sub-wavelength outcoupling facet, we can still observe a change in far-field after implementing the side-absorber. Figure\,\ref{fig:farfield} displays the measured far-field before (a) and after (b) the fabrication of the side-absorber. It is evident that due to the absence of $\text{TM}_{01}$ modes the beam pattern of the laser with a side-absorber is more directional in the $\phi$-angle. This is explained by the fact that the far-field of $\text{TM}_{00}$ modes is more directional as the one of $\text{TM}_{01}$ modes \cite{Gellie2008}. Therefore, such a change in far-field can be expected. As the detector used for the THz power measurements in Fig.\,\ref{fig:sem-cc}(b) covers a limited solid angle only, more power is detected in the case of a more directional far-field. Thus, the total laser power is not necessarily increased by the usage of side-absorbers.

Having proven the successful suppression of higher order lateral modes in standard THz QCLs and knowing the right amount of set-back, either from simulations or experiments, we are now able to directly fabricate a laser with the suitable set-back using a SiNx hard mask for the dry etching step, including an emitter section for the THz pulse generation (as described in section \ref{sec:fabrication}). Figure\,\ref{fig:spectrum-setback}(c) shows a laser spectrum where the set-back has been implemented directly and the device includes an emitter section, as shown in Fig.\,\ref{fig:sem-cc}(a). Comparing Fig.\,\ref{fig:spectrum-setback}(b) and (c) emphasizes that the two different fabrication procedures (direct and post-processing) of the side-absorbers lead to very similar results. The smaller mode spacing in Fig.\,\ref{fig:spectrum-setback}(c) is due to the longer cavity of this laser (2.93\,mm instead of 2\,mm). The modes at 2.64 and 2.78\,THz are attenuated by residual water vapor (black line), since this spectrum was measured in a nitrogen-purged Fourier transform infrared spectroscopy (FTIR) system while the other spectra were measured using a vacuum-pumped FTIR. This also explains the lower signal-to-noise ratio (SNR) in Fig.\,\ref{fig:spectrum-setback}(c), resulting in a smaller bandwidth above the noise floor.

\section{Pulse generation and amplification}
Apart from using THz QC heterostructures to build lasers, they are obvious candidates for the generation and amplification of ultra-short pulses due to their broad bandwidth. The requirements for coherent and phase-resolved detection of THz QCL radiation can be provided by THz time-domain spectroscopy (TDS) systems \cite{kroell2007}. Hence, it was shown that THz QCLs can emit a stable train of pulses by injecting seeding with a weak THz pulse \cite{Maysonnave2012}, by active mode-locking \cite{barbieri2011}, or a combination of both \cite{Freeman2013}. In all three cases a bound-to-continuum (BTC) based QCL fabricated with a single-plasmon waveguide has been used. The generated pulses exhibited lengths (full width at half-maximum - FWHM) of 9\,ps (11 lasing modes within $\sim$\,200\,GHz) \cite{Maysonnave2012}, 10\,ps (10 lasing modes within $\sim$\,150\,GHz) \cite{barbieri2011}, and 15\,ps (6 lasing modes within $\sim$\,100\,GHz) \cite{Freeman2013}, respectively. However, achieving pulse lengths in the order of 1\,ps requires a bandwidth of $\sim$1\,THz. It is therefore essential to use heterogeneous active regions based on inherently broadband resonant longitudinal optical phonon depopulation (RP) designs, as the one used in this work.

Using the coupled cavity device geometry, as discussed in section \ref{sec:fabrication} (without side-absorbers), and a similar version of the heterogeneous active region presented in this work, we were previously able to achieve broadband amplification over a bandwidth of 500\,GHz and amplification factors exceeding 20\,dB \cite{bachmann2015}. The amplification process relies on gain switching that is initiated by a RF pulse. Injection seeding permits the coherent detection of the THz electric field pulse train, separated by the round-trip time of the QCL cavity \cite{kroell2007,oustinov2010}. After a few round-trips, the individual pulses merge and a rich sub-pulse fine structure is formed \cite{bachmann2015}. Its presence can be explained by higher order lateral modes that are seeded by the integrated THz emitter. Different lateral modes have the tendency to form separate pulses, propagating with the group velocities of the given mode order. Therefore, the presented mode selection helps to improve the THz amplifier performance.

To investigate the influence of lossy side-absorbers on the pulse formation dynamics we performed injection seeding experiments. They are based on the ones presented in Ref.\,\cite{bachmann2015}, but with the optimized active region reported in Ref.\,\cite{Roesch2014}. The seeding experimental setup features a Ti:Sapphire laser based THz-TDS system adapted to the coupled cavity geometry of our samples \cite{Martl2011,Bachmann2014}. Near-infrared femtosecond (fs) laser pulses (790\,nm, $\sim$\,80\,fs) are used to generate broadband THz seeding pulses in the emitter section, which are directly coupled into the amplifier section. A hyper-hemispherical silicon lens attached to the THz amplifier output facet is used to improve the collection efficiency and achieve a better THz focus on the 1\,mm thick ZnTe detection crystal. The measurements were done at a heat-sink temperature of 10\,K.

\begin{figure}[htbp]
	\centering
	\includegraphics[width=0.7\linewidth]{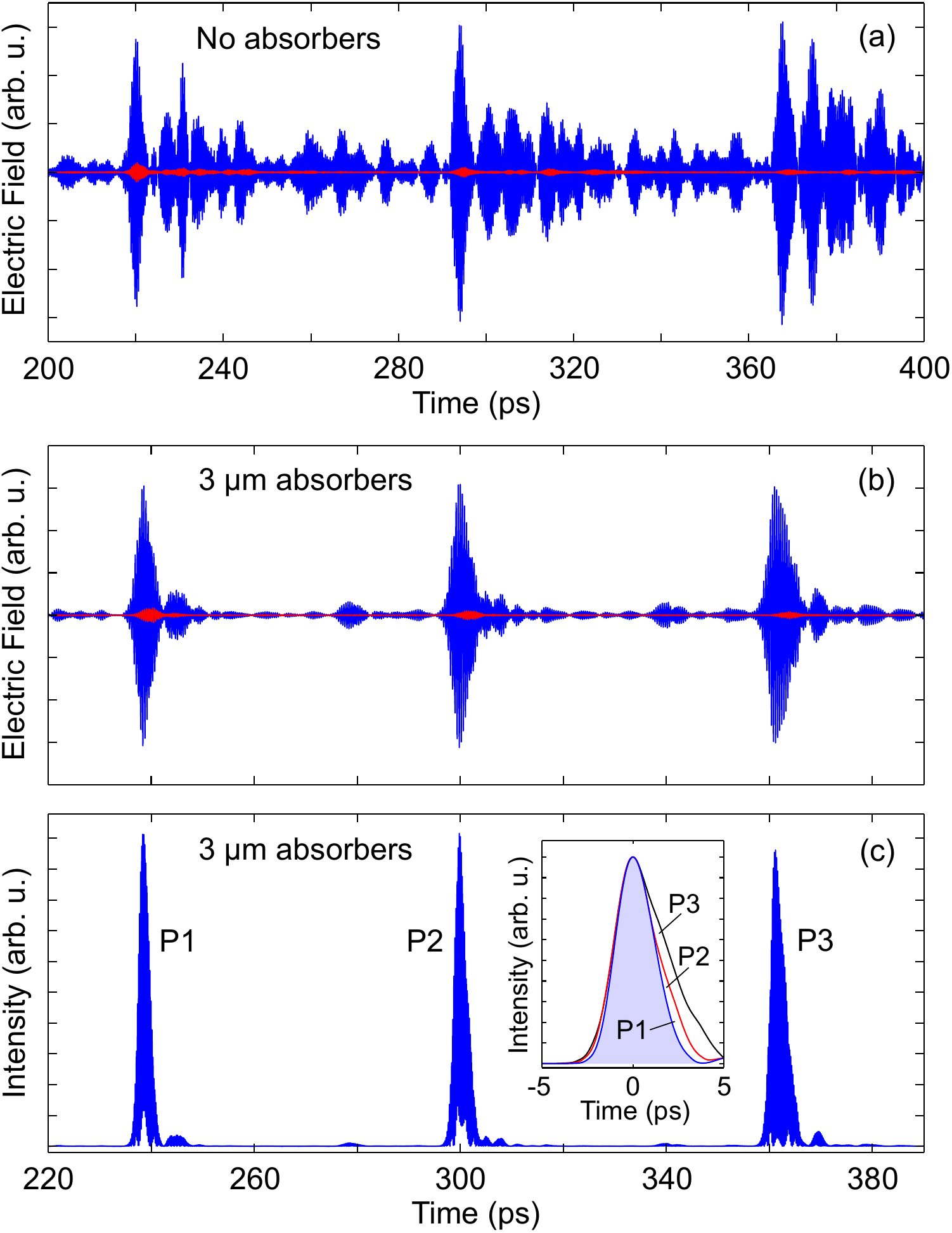}
	\caption{Injection seeded THz emission of coupled cavity metal-metal QCLs, operated in the AMP (blue lines) and the REF mode (red lines). (a) Standard device without lossy side-absorbers. (b) Mode controlled device with side-absorbers. (c)\,Squared THz electric field pulse train of (b). The inset shows the intensity profiles of the three pulses, revealing a pulse length of 2.5\,ps for the first pulse (P1).}
	\label{fig:pulse-train}
\end{figure}

The seeding experiment is done for two different operating modes of the QC amplifier section. In the reference mode (REF), the amplifier section is driven just above threshold in order to guarantee transparency conditions. In the amplification mode (AMP), the amplifier section is driven $\sim$\,20\% below threshold and a sub-nanosecond long RF pulse with a peak power of +28\,dBm is used to switch the gain while a THz seed pulse is injected into its cavity. First, we have investigated a coupled cavity device without lossy side-absorbers and the dimensions of 2.96\,mm\,x\,70\,$\mu$m. Figure\,\ref{fig:pulse-train}(a) shows the injection seeded part of the QC amplifier THz electric field output (AMP mode, blue line) along with the decreasing pulse train amplitudes of the REF mode (red line). The generated pulses are spaced by the round-trip time through the amplifier waveguide. In agreement with our previous experiments, the coherent QCL emission does not exhibit a train of distinct round-trip pulses \cite{bachmann2015}, but a very complex and highly dynamic fine structure on a ps time scale is formed. Similar behavior was observed in a homogeneous RP active region metal-metal THz QCL \cite{Wang2015}.

In contrast to that, the results of the seeding experiment change dramatically if a coupled cavity device of comparable size (2.48\,mm\,x\,75\,$\mu$m), but with side-absorbers (see Fig.\,\ref{fig:sem-cc}(a)), is used. As can be seen in Fig.\,\ref{fig:pulse-train}(b), if the THz amplifier is driven in the AMP mode (blue line), a train of clearly separated THz pulses is formed. A comparison of the two waveguide configurations, seen in Fig.\,\ref{fig:pulse-train}(a) and (b), clearly demonstrates the influence of higher order lateral modes on QC heterostructure based THz pulse generation. The significantly lower effective refractive index of the $\text{TM}_{01}$ with respect to the $\text{TM}_{00}$ modes leads to the formation of sub-pulses due to different group velocities and no clean THz pulse train can be formed. The parasitic pulses appearing $\sim$\,22\,ps before the main pulses are residues of an air-side mode that is bound to the metal-air interface \cite{Martl2011} and result in an interleaved pulse train besides the desired waveguide mode. Figure\,\ref{fig:pulse-train}(c) displays the squared electric field pulse train of Fig.\,\ref{fig:pulse-train}(b) and the inset shows the temporal profiles of the three intensity pulses in the main panel. The pulse centered at 239\,ps (P1) exhibits a Gaussian like pulse shape and a record ultra-short pulse length of 2.5\,ps (FWHM). The increasing pulse widths and the varying tails of the subsequent pulses indicate significant higher order dispersion of the $\text{TM}_{00}$ modes in the laser cavity.

\begin{figure}[htbp]
	\centering
	\includegraphics[width=0.7\linewidth]{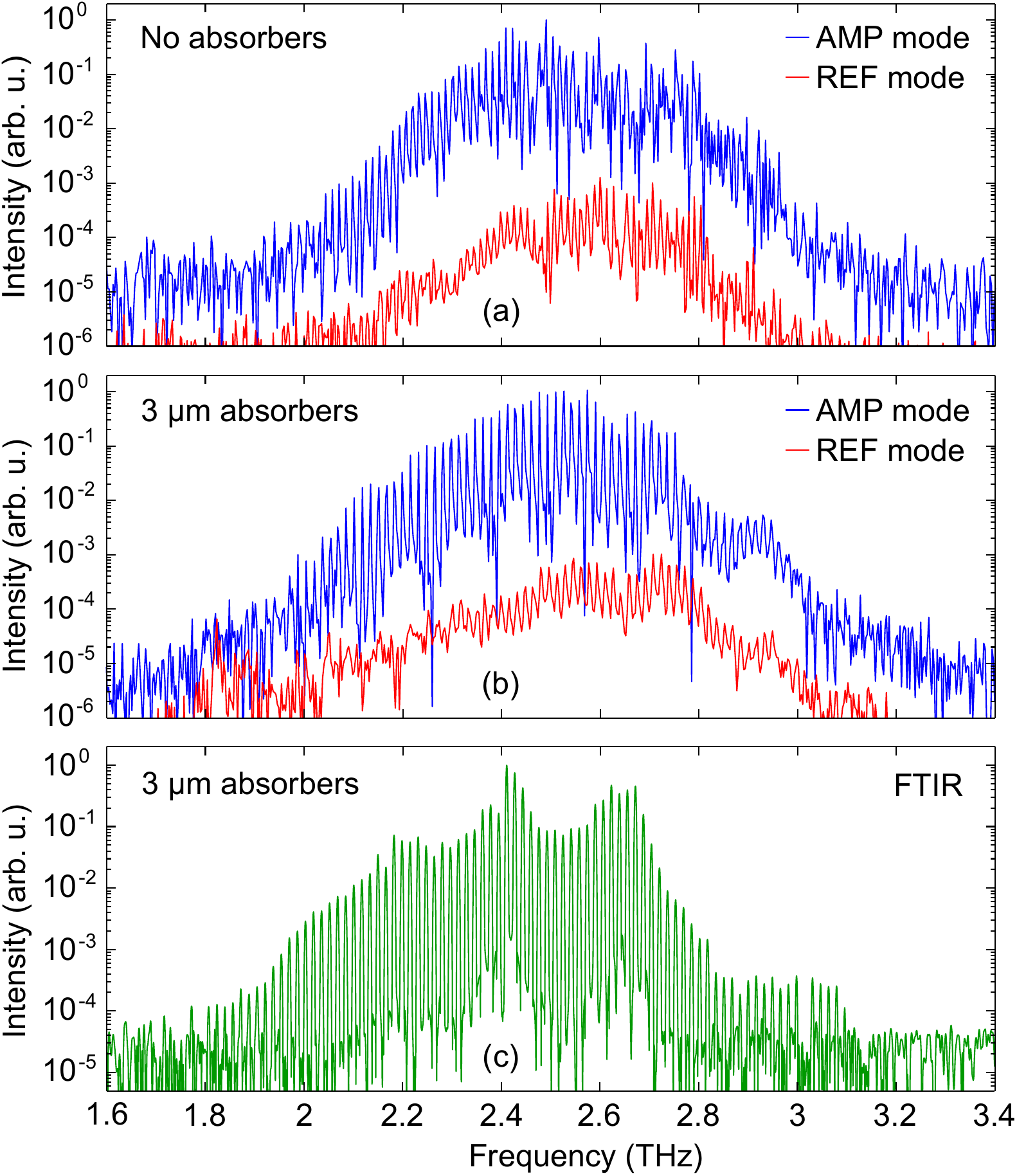}
	\caption{Fourier transformed TDS traces of a standard (a) and a spectrally optimized (b) heterogeneous QC amplifier device, driven in the AMP (blue lines) and REF (red lines) mode. (c) Non-seeded laser spectrum (FTIR) of the optimized device from (b) at the NDR point.}
	\label{fig:spectrum-tds}
\end{figure}

In order to check the seeded QCL bandwidth and the influence of excited lateral modes on the spectral properties, 360\,ps long versions of the TDS time traces shown in Fig.\,\ref{fig:pulse-train}(a),\,(b) were Fourier transformed to the frequency domain. The corresponding intensity spectra are shown in Fig.\,\ref{fig:spectrum-tds}(a),\,(b). Independent of side-absorber usage, the two AMP spectra verify that all three active regions, centered at 2.3., 2.6 and 2.9 THz, are seeded and demonstrate a SNR of 50\,dB. Comparing the AMP with the REF spectra reveals amplification bandwidths exceeding 1\,THz and amplification factors of $\sim$\,30\,dB. In addition, the seeded spectra display a significantly improved power distribution, compared to our previous results \cite{bachmann2015}, with a fairly flat top between 2.1 and 2.8\,THz (20\,dB intensity range). The device without side-absorbers, shown in Fig.\,\ref{fig:spectrum-tds}(a), displays a spectrum with non-uniform mode spacing including higher order lateral modes, especially for frequencies above 2.6\,THz. Similar irregularities in the seeded TDS spectrum of a 60\,$\mu$m wide RP based metal-metal QCL with an integrated planar horn antenna were also seen in Ref.\,\cite{wang2016}. On the contrary, and as expected from the clean pulse train in Fig.\,\ref{fig:pulse-train}(b), the spectrum of the QC amplifier with lossy side-absorbers, shown in Fig.\,\ref{fig:spectrum-tds}(b), exhibits a constant mode spacing over the entire bandwidth with a total number of 60 fundamental Fabry-Perot modes. Assuming a Gaussian intensity profile for the pulses in Fig.\,\ref{fig:pulse-train}(c) and using a transform-limited pulse length of 2.5\,ps leads to a corresponding bandwidth of 180\,GHz. This agrees very well with the measured FWHM bandwidth in Fig.\,\ref{fig:spectrum-tds}(b).

For comparison, Fig.\,\ref{fig:spectrum-tds}(c) displays a non-seeded spectrum of the spectrally optimized device from Fig.\,\ref{fig:spectrum-tds}(b). The laser is operated at the NDR point and acquired by a FTIR spectrometer with a spectral resolution of 2.25\,GHz. The difference of the spectra in Fig.\,\ref{fig:spectrum-tds}(b) and \ref{fig:spectrum-tds}(c) at frequencies above 2.8\,THz are possibly related to higher order dispersion in this part of the spectrum, affecting the pulse propagation as visible in the inset of Fig.\,\ref{fig:pulse-train}(c).

\section{Discussion and Conclusion}
The suppression of any kind of higher order lateral modes in broadband THz QC lasers is an essential requirement for many applications, especially for frequency comb operation and short pulse generation. To increase the waveguide losses of higher order lateral modes and thereby preventing them from lasing, we added side-absorbers to the metal-metal resonators. The reported fabrication technique, using a focused ion beam to fabricate set-backs in addition with the deposition of a several nm thick nickel layer, enables a convenient implementation of side-absorbers onto processed and functioning lasers.

Frequency dependent threshold gain calculations were used to verify the broadband mode suppression mechanism and to adapt the total amount of side-absorber losses with respect to the used laser ridge width. Several devices, with different ridge and side-absorber widths, have been fabricated and examined, demonstrating the versatility and reliability of the presented side-absorber technique. The performed simulations explain qualitatively very well the experimental results. 

The optimized devices show significantly improved spectral properties in terms of lasing bandwidth and uniformity of the power distribution with a flat top exceeding 700\,GHz within a 10\,dB range. This is attributed to the suppressed gain competition of fundamental and higher order lateral modes. Lasers with side-absorbers also exhibit an increase in output power of about 30\%, which is mainly due to a more directional far-field and a finite collection solid angle of the optical power measurement.

Implementing side-absorbers on injection seeded THz QCLs leads to a manifold improvement of the pulse generation properties, resulting in the formation of a clean train of transform-limited THz pulses with a record ultra-short pulse length of 2.5\,ps. Together with the seeded bandwidth of $\sim$\,1\,THz, this makes such devices very useful as powerful sources for THz-TDS systems, especially for frequencies above 2\,THz, where the power and SNR of conventional TDS systems rapidly decreases. 

The presented mode control of broadband THz QCLs supports the development of octave-spanning frequency combs with flat topped lasing spectra and the generation of sub-picosecond long THz pulses.

\section*{Funding Information}
The authors acknowledge partial financial support by the European Union via the FET-Open grant TERACOMB (ICT-296500), the Austrian Science Fund (FWF) through the SFB NextLite (F4902), the DK Solids4Fun (W1243) and the Swiss National Science Foundation (SNF) through project 200020 165639.

\section*{Acknowledgments}
The authors acknowledge the Center for Micro- and Nanostructures (ZMNS) at TU Wien and the joint labs FIRST and ScopeM at ETH Zurich for sample processing. We further thank C. Bonzon for the support on the 2D simulations, K. Otani for the support and discussions concerning the device fabrication and C. Derntl for the support with the SEM.

\bibliographystyle{unsrt}

\begin{thebibliography}{10}
	\newcommand{\enquote}[1]{``#1''}
	
	\bibitem{faist1994}
	J.~Faist, F.~Capasso, D.~L. Sivco, C.~Sirtori, A.~L. Hutchinson, and A.~Y. Cho,
	\enquote{{Quantum Cascade Laser},} Science \textbf{264}, 553--556 (1994).
	
	\bibitem{turcinkova2011}
	D.~Tur\v{c}inkov{\'a}, G.~Scalari, F.~Castellano, M.~I. Amanti, M.~Beck, and
	J.~Faist, \enquote{{Ultra-broadband heterogeneous quantum cascade laser
			emitting from 2.2 to 3.2 THz},} Appl. Phys. Lett. \textbf{99}, 191104 (2011).
	
	\bibitem{Roesch2014}
	M.~R{\"o}sch, G.~Scalari, M.~Beck, and J.~Faist, \enquote{{Octave-spanning
			semiconductor laser},} Nat. Photonics \textbf{9}, 42--47 (2015).
	
	\bibitem{Gmachl2002}
	C.~Gmachl, D.~L. Sivco, R.~Colombelli, F.~Capasso, and A.~Y. Cho,
	\enquote{{Ultra-broadband semiconductor laser.}} {Nature} \textbf{415},
	883--887 (2002).
	
	\bibitem{Roesch2016}
	M.~R{\"o}sch, G.~Scalari, G.~Villares, L.~Bosco, M.~Beck, and J.~Faist,
	\enquote{{On-chip, self-detected terahertz dual-comb source},} Appl. Phys.
	Lett. \textbf{108}, 171104 (2016).
	
	\bibitem{Burghoff2014}
	D.~Burghoff, T.-Y. Kao, N.~Han, I.~C.~C. Wang, X.~Cai, Y.~Yang, D.~J. Hayton,
	J.-R. Gao, J.~L. Reno, and Q.~Hu, \enquote{{Terahertz laser frequency
			combs},} Nat. Photonics \textbf{8}, 462--467 (2014).
	
	\bibitem{bachmann2015}
	D.~Bachmann, N.~Leder, M.~R{\"o}sch, G.~Scalari, M.~Beck, H.~Arthaber,
	J.~Faist, K.~Unterrainer, and J.~Darmo, \enquote{{Broadband terahertz
			amplification in a heterogeneous quantum cascade laser},} Opt. Express
	\textbf{23}, 3117--3125 (2015).
	
	\bibitem{oustinov2010}
	D.~Oustinov, N.~Jukam, R.~Rungsawang, J.~Mad{\'e}o, S.~Barbieri, P.~Filloux,
	C.~Sirtori, X.~Marcadet, J.~Tignon, and S.~Dhillon, \enquote{{Phase seeding
			of a terahertz quantum cascade laser},} Nat. Commun. \textbf{1}, 1--6 (2010).
	
	\bibitem{barbieri2011}
	S.~Barbieri, M.~Ravaro, P.~Gellie, G.~Santarelli, C.~Manquest, C.~Sirtori,
	S.~P. Khanna, E.~H. Linfield, and A.~G. Davies, \enquote{{Coherent sampling
			of active mode-locked terahertz quantum cascade lasers and frequency
			synthesis},} Nat. Photonics \textbf{5}, 306--313 (2011).
	
	\bibitem{Freeman2013}
	J.~R. Freeman, J.~Maysonnave, H.~E. Beere, D.~A. Ritchie, J.~Tignon, and S.~S.
	Dhillon, \enquote{{Electric field sampling of modelocked pulses from a
			quantum cascade laser},} Opt. Express \textbf{21}, 16162--16169 (2013).
	
	\bibitem{Wang2015}
	F.~Wang, K.~Maussang, S.~Moumdji, R.~Colombelli, J.~R. Freeman, I.~Kundu,
	L.~Li, E.~H. Linfield, A.~G. Davies, J.~Mangeney, J.~Tignon, and S.~S.
	Dhillon, \enquote{{Generating ultrafast pulses of light from quantum cascade
			lasers},} Optica \textbf{2}, 944--949 (2015).
	
	\bibitem{williams2003}
	B.~S. Williams, S.~Kumar, H.~Callebaut, Q.~Hu, and J.~L. Reno,
	\enquote{{Terahertz quantum-cascade laser at 100 $\mu$m using metal waveguide
			for mode confinement},} Appl. Phys. Lett. \textbf{83}, 2124--2126 (2003).
	
	\bibitem{scalari2008review}
	G.~Scalari, C.~Walther, M.~Fischer, R.~Terazzi, H.~Beere, D.~Ritchie, and
	J.~Faist, \enquote{{THz and sub-THz quantum cascade lasers},} Laser and
	Photonics Reviews \textbf{3}, 45--66 (2009).
	
	\bibitem{kohen2005}
	S.~Kohen, B.~S. Williams, and Q.~Hu, \enquote{{Electromagnetic modeling of
			terahertz quantum cascade laser waveguides and resonators},} J. Appl. Phys.
	\textbf{97}, 053106 (2005).
	
	\bibitem{williams2007}
	B.~S. Williams, \enquote{{Terahertz quantum-cascade lasers},} Nat. Photonics
	\textbf{1}, 517--525 (2007).
	
	\bibitem{Adam2006}
	A.~J.~L. Adam, I.~Ka\v{s}alynas, J.~N. Hovenier, T.~O. Klaassen, J.~R. Gao,
	E.~E. Orlova, B.~S. Williams, S.~Kumar, Q.~Hu, and J.~L. Reno, \enquote{{Beam
			patterns of terahertz quantum cascade lasers with subwavelength cavity
			dimensions},} Appl. Phys. Lett. \textbf{88}, 151105 (2006).
	
	\bibitem{Orlova2006}
	E.~E. Orlova, J.~N. Hovenier, T.~O. Klaassen, I.~Ka\v{s}alynas, A.~J.~L. Adam,
	J.~R. Gao, T.~M. Klapwijk, B.~S. Williams, S.~Kumar, Q.~Hu, and J.~L. Reno,
	\enquote{{Antenna Model for Wire Lasers},} Phys. Rev. Lett. \textbf{96},
	173904 (2006).
	
	\bibitem{KOE02}
	R.~K{\"o}hler, A.~Tredicucci, F.~Beltram, H.~E. Beere, E.~H. Linfield, A.~G.
	Davies, D.~A. Ritchie, R.~C. Iotti, and F.~Rossi, \enquote{{Terahertz
			Heterostructure Laser},} Nature \textbf{417}, 156 (2002).
	
	\bibitem{Maineult2008}
	W.~Maineult, P.~Gellie, A.~Andronico, P.~Filloux, G.~Leo, C.~Sirtori,
	S.~Barbieri, E.~Peytavit, T.~Akalin, J.-F. Lampin, H.~E. Beere, and D.~A.
	Ritchie, \enquote{{Metal-metal terahertz quantum cascade laser with
			micro-transverse-electromagnetic-horn antenna},} Appl. Phys. Lett.
	\textbf{93}, 183508 (2008).
	
	\bibitem{Fan2008}
	J.~A. Fan, M.~A. Belkin, F.~Capasso, S.~P. Khanna, M.~Lachab, A.~G. Davies, and
	E.~H. Linfield, \enquote{{Wide-ridge metal-metal terahertz quantum cascade
			lasers with high-order lateral mode suppression},} Appl. Phys. Lett.
	\textbf{92}, 031106 (2008).
	
	\bibitem{xu2013}
	C.~Xu, S.~G. Razavipour, Z.~Wasilewski, and D.~Ban, \enquote{{An investigation
			on optimum ridge width and exposed side strips width of terahertz quantum
			cascade lasers with metal-metal waveguides},} Opt. Express \textbf{21},
	31951--31959 (2013).
	
	\bibitem{Tanvir2013}
	H.~Tanvir, B.~M.~A. Rahman, and K.~T.~V. Grattan, \enquote{{A Higher Order
			Lateral Mode Suppression Scheme for Terahertz Quantum Cascade Laser
			Waveguides},} IEEE J. Sel. Top. Quantum Electron. \textbf{19}, 8501106
	(2013).
	
	\bibitem{Gellie2008}
	P.~Gellie, W.~Maineult, A.~Andronico, G.~Leo, C.~Sirtori, S.~Barbieri,
	Y.~Chassagneux, J.~R. Coudevylle, R.~Colombelli, S.~P. Khanna, E.~H.
	Linfield, and A.~G. Davies, \enquote{{Effect of transverse mode structure on
			the far field pattern of metal-metal terahertz quantum cascade lasers},} J.
	Appl. Phys. \textbf{104}, 124513 (2008).
	
	\bibitem{faist2013book}
	J.~Faist, \emph{{Quantum Cascade Lasers}} (Oxford University Press, 2013).
	
	\bibitem{saleh_teich_book}
	B.~Saleh and M.~Teich, \emph{{Fundamentals of Photonics }} (Wiley Series in
	Pure and Applied Optics, 2007), 2nd ed.
	
	\bibitem{Martl2011}
	M.~Martl, J.~Darmo, C.~Deutsch, M.~Brandstetter, A.~M. Andrews, P.~Klang,
	G.~Strasser, and K.~Unterrainer, \enquote{{Gain and losses in THz quantum
			cascade laser with metal-metal waveguide},} Opt. Express \textbf{19},
	733--738 (2011).
	
	\bibitem{Bachmann2014}
	D.~Bachmann, M.~R{\"o}sch, C.~Deutsch, M.~Krall, G.~Scalari, M.~Beck, J.~Faist,
	K.~Unterrainer, and J.~Darmo, \enquote{{Spectral gain profile of a
			multi-stack terahertz quantum cascade laser},} Appl. Phys. Lett.
	\textbf{105}, 181118 (2014).
	
	\bibitem{wang2016}
	F.~Wang, I.~Kundu, L.~Chen, L.~Li, E.~H. Linfield, A.~G. Davies, S.~Moumdji,
	R.~Colombelli, J.~Mangeney, J.~Tignon, and S.~S. Dhillon,
	\enquote{{Engineered far-fields of metal-metal terahertz quantum cascade
			lasers with integrated planar horn structures},} Opt. Express \textbf{24},
	2174--2182 (2016).
	
	\bibitem{hugi2012}
	A.~Hugi, G.~Villares, S.~Blaser, H.~C. Liu, and J.~Faist,
	\enquote{{Mid-infrared frequency comb based on a quantum cascade laser.}}
	{Nature} \textbf{492}, 229--233 (2012).
	
	\bibitem{kroell2007}
	J.~Kr{\"o}ll, J.~Darmo, S.~S. Dhillon, X.~Marcadet, M.~Calligaro, C.~Sirtori,
	and K.~Unterrainer, \enquote{{Phase-resolved measurements of stimulated
			emission in a laser},} Nature \textbf{449}, 698--701 (2007).
	
	\bibitem{Maysonnave2012}
	J.~Maysonnave, K.~Maussang, J.~R. Freeman, N.~Jukam, J.~Madeo, P.~Cavalie,
	R.~Rungsawang, S.~P. Khanna, E.~H. Linfield, A.~G. Davies, H.~E. Beere, D.~A.
	Ritchie, S.~S. Dhillon, and J.~Tignon, \enquote{{Mode-locking of a terahertz
			laser by direct phase synchronization},} Opt. Express \textbf{20},
	20855--20862 (2012).	
\end{thebibliography}

\newpage

\vspace{1cm}

\end{document}